\documentclass[journal]{IEEEtran}

\usepackage{amsmath,amsfonts}
\usepackage{array}
\usepackage{textcomp}
\usepackage{stfloats}
\usepackage{url}
\usepackage{verbatim}
\usepackage{graphicx}
\usepackage{cite}
\usepackage{amssymb}
\usepackage{mathtools}
\usepackage[colorlinks,linkcolor=black,anchorcolor=black,citecolor=black]{hyperref}

\usepackage[]{footmisc}  

\usepackage[ruled,linesnumbered]{algorithm2e}
\SetKw{Continue}{continue}
\SetKw{Break}{break}
\SetKw{And}{and}
\SetKw{From}{from}
\SetKw{To}{to}
\SetKw{Step}{step}

\usepackage{algorithmic}
\usepackage{booktabs}
\usepackage{listings}
\usepackage{multirow}

\usepackage{multirow}
\newcolumntype{P}[1]{>{\centering\arraybackslash}p{#1}}
\newcolumntype{M}[1]{>{\centering\arraybackslash}m{#1}}

\UseRawInputEncoding

\usepackage{xcolor} 

\begin{document}

\title{PAC Code Rate-Profile Design Using Search-Constrained Optimization Algorithms}
\author{Mohsen~Moradi\textsuperscript{\href{https://orcid.org/0000-0001-7026-0682}{\includegraphics[scale=0.06]{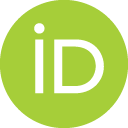}}} 
and
David~G.~M.~Mitchell\textsuperscript{\href{https://orcid.org/0000-0002-3544-9225}{ \includegraphics[scale=0.06]{figs/ORCID}}}
\thanks{The authors are with the Klipsch School of Electrical and Computer Engineering, New Mexico State University, Las Cruces, NM 88003, USA (e-mail: moradi23@nmsu.edu, dgmm@nmsu.edu).}
}

\maketitle
\begin{abstract}
In this paper, we introduce a novel rate-profile design based on search-constrained optimization techniques to assess the performance of polarization-adjusted convolutional (PAC) codes under Fano (sequential) decoding. 
The results demonstrate that the resulting PAC code offers much reduced computational complexity compared to a construction based on a conventional genetic algorithm without a performance loss in error-correction performance.
As the fitness function of our algorithm, we propose an adaptive successive cancellation list decoding algorithm to determine the weight distribution of the rate profiles.
The simulation results indicate that, for a PAC(256, 128) code,
only 8$\%$ of the population requires that their fitness function be evaluated with a large list size.
This represents an improvement of almost 92$\%$ over a conventional evolutionary algorithm.
For a PAC(64, 32) code, this improvement is about 99$\%$.
We also plotted the performance of the high-rate PAC(128, 105) and PAC(64, 51) codes, and the results show that they exhibit superior performance compared to other algorithms.
\end{abstract}

\section{Introduction}
Recently, Ar{\i}kan presented polarization-adjusted convolutional (PAC) codes that can achieve the theoretical bounds \cite{polyanskiy2010channel} with Reed-Muller (RM) rate profiling at a block length of $N = 128$.
This construction has better error-correction performance than the 5G polar codes \cite{arikan2019sequential, moradi2020performance}. 
PAC codes combine polar and convolutional codes, and its encoder may be seen as a tree code, enabling the use of a tree search decoding algorithms such as Fano \cite{fano1963heuristic} or successive cancellation list (SCL) \cite{tal2015list} decoding.
The minimum distance and weight distribution of the PAC and pre-transformed and post-transformed polar codes have a significant impact on the error-correction performance of the codes and have been actively studied in recent years
\cite{li2019pre, moradi2023minimum, yao2023deterministic}.

An RM code-based design for the rate profile of PAC codes provides a good weight distribution, resulting in PAC codes with excellent error-correction performance.  
Note, however, that only a limited number of code rates are available for the RM code; therefore, attaining a PAC code with an arbitrary rate necessitates the use of other rate profile construction techniques. 
Furthermore, for the existing RM-based code construction of PAC codes with moderate block lengths of $N=256$, the computational complexity can also be quite high \cite{moradi2023application}.
In \cite{tonnellier2021systematic}, numerical findings demonstrate that a PAC code is also capable of achieving the theoretical finite length limit for a block length of $N=256$ by using a genetic algorithm to determine the code's rate profile. 
The objective of the optimization technique in that paper is to improve an estimate of the upper bound of the code's error-correction performance. 
However, even if the resulting code has high reliability, ignoring the computational complexity of decoding during the design process can result in a significant increase in decoder latency. 
Using a heuristic technique for the construction of the rate profile, it was shown in \cite{moradi2021monte} that PAC codes can approach near the theoretical bounds for block lengths of $N=64$ and $N=256$ with much lower computational complexity than the construction obtained using the genetic algorithm. 
\IEEEpubidadjcol

Considering the polarization of the cutoff rate \cite{moradi2021monte}, a weight sum-based metric for rate profile construction for PAC codes is proposed in \cite{liu2022weighted}. 
In \cite{moradi2023application}, by freezing a few data bits, a technique to determine the rate profile based on the polarization of the computational cutoff rate is presented. 
As a result, sequential decoding of the resulting PAC code has much lower computational complexity than that of the RM rate profile. 
This code construction imposes a restriction on the data bit positions of the rate profile.
Similar to the RM rate profile, this approach generates a code with a limited amount of code rates. 
The most important takeaway from this study is that the approach establishes a code rate profile constraint that has to be fulfilled for low-computational complexity decoding of a code.

In this paper, we present a novel evolutionary based rate-profile design method.
Our proposed algorithm is able to approach the theoretical error-correction bounds while maintaining a reasonable decoding computational complexity.
Moreover, in our proposed algorithm, we generalize the code construction of \cite{moradi2023application} to any code rate.
We propose an adaptive successive cancellation list (SCL) decoding algorithm to evaluate the fitness function. Simulation results demonstrate that, when attempting to obtain an optimal rate profile for a PAC$(256, 128)$ code, our algorithm requires approximately $92\%$ fewer attempts to evaluate the fitness function with a large list size than a conventional evolutionary algorithm.
The empirical results demonstrate that the proposed rate profile design yields error-correction performance comparable to that of state-of-the-art codes.
In addition, the constraint placed on the search space accelerates the convergence of the search algorithm. 
Our results outperform \cite{liu2022weighted} in terms of both error-correction performance and computational complexity.

The proposed algorithm identifies a rate profile with an error-correction performance comparable to the one in \cite{tonnellier2021systematic}, and the computational complexity of the Fano algorithm for the rate profile that we obtained is more than 10 times less than that paper.
This paper demonstrates the importance of the polarization involved in the rate profile construction of pre-transformed polar codes, whereas \cite{arikan2006channel} investigates the polarization of the cutoff rate and \cite{moradi2023application} investigates the polarization of the computation required for sequential decoding of pre-transformed polar codes.


Throughout, vectors and matrices are specified with bold text, and all operations are performed on a binary field.
For a vector $\mathbf{u} = (u_1, u_2, \ldots, u_N)$  we use $\mathbf{u}^i$ to represent subvector $(u_1, \ldots, u_i)$ and $\mathbf{u}_i^j$ to express subvector $(u_i, \ldots, u_j)$.

\section{PAC Coding} \label{sec: scheme}

A PAC code with the code block length of $N = 2^n$ and data length of $K$, denoted PAC$(N, K)$, can be specified by its rate profile $\mathcal{A}$ and a connection polynomial $\mathbf{c}(x) = c_0 + c_1x + \cdots +  c_mx^{m}$ s.t. $c_0 = c_m = 1$.
For encoding, first, the length $K$ information vector $\mathbf{d}$ would be inserted into the length $N$ vector $\mathbf{v}$ as $\mathbf{v}_{\mathcal{A}} = \mathbf{d}$ and $\mathbf{v}_{\mathcal{A}^{c}} = \mathbf{0}$, where the subscripts indicate a set of vector indices corresponding to the location of respective data, and $\mathcal{A}^{c}$ is the complement of the set $\mathcal{A}$.
Then, the convolutional encoder encodes the data carrier vector $\mathbf{v}$ as $\mathbf{u} = \mathbf{v}\mathbf{T}$ and the polar encoder subsequently encodes it as $\mathbf{x} = \mathbf{u} \mathbf{F}^{\otimes n}$, where 
the matrix $\mathbf{T}$ is an upper triangular Toeplitz matrix constructed by the coefficients of the polynomial $\mathbf{c}(x)$,
and $\mathbf{F}^{\otimes n}$ is the $n$th Kronecker power of the matrix $\mathbf{F} = \begin{bsmallmatrix} 1 & 0\\ 1 & 1 \end{bsmallmatrix}$. 
The codeword $\mathbf{x}$ is transmitted over a binary input additive white Gaussian noise (BI-AWGN) channel, and $\mathbf{y}$ is received as the channel output.

\section{Sequential decoding and its complexity} \label{sec: sequential decoding}

At the receiver, a sequential decoder generates an estimate $\hat{\mathbf{v}}$ of the data carrier vector $\mathbf{v}$ with the assistance of a metric calculator and then extracts an estimate of the data word by setting $\hat{\mathbf{d}} = \hat{\mathbf{v}}_\mathcal{A}$. 
The frame error rate (FER), which is defined as the probability $\text{P}(\mathbf{d} \neq \hat{\mathbf{d}})$, is the primary performance metric used to assess the code. 

For a sequential decoder, we implement the Fano algorithm introduced in \cite{moradi2021sequential} with the polarized cutoff rates as the bias values and the connection polynomial $\mathbf{c}(x) = 1+x^3+x^7+x^9+x^{10}$.
The Fano decoder searches the coding tree for the correct path according to the PAC code. 
To measure the decoding complexity, we use the average number of visits (ANV) of the Fano decoder per decoded bit \cite{moradi2021sequential}.

For convolutional codes, the cutoff rate defines the range of rates for which the average computational complexity of sequential decoding is finite or infinite. 
In particular, it has been shown that the average number of guesses necessary to decode a codeword is bounded as \cite{arikan1996inequality}
\begin{equation} 
    \mathbb{E}[G_{N}]  \geq e^{N(R-R_0(W) - o(N) )}.
\end{equation}
This indicates that, at rates beyond the cutoff rate, the computational complexity of sequential decoding increases exponentially with the code block length. 

Define $K^-$ as the number of data bits in the vector $\mathbf{v}^{N/2}$ (referred to as the first part) and $K^{+}$ as the number of data bits in the vector $\mathbf{v}_{N/2 +1}^{N}$ (the second part); hence, $K = K^- + K^+$.
Define $R_0(W^-)$ and $R_0(W^+)$ as the polarized channel cutoff rates after the first step of polarization and $R^-$ and $R^+$ as 
    $R^- \triangleq \frac{K^-}{N/2}, ~~~ R^+ \triangleq \frac{K^+}{N/2};$
hence, $R^- + R^+ = 2R$.
The complexity bounds for the first and second parts (corresponding to the $W^-$ and $W^+$ synthesized channels) are a function of $N/2$, given by 
\begin{equation}
    \mathbb{E}[G_{N/2}] \geq e^{\frac{N}{2}(R^{-}-R_0(W^-) - o(\frac{N}{2}))}
\end{equation}
and
\begin{equation}
    \mathbb{E}[G_{N/2,N}] \geq e^{\frac{N}{2}(R^{+}-R_0(W^+) - o(\frac{N}{2}) )}.
\end{equation}
This indicates that the computational cutoff rate of sequential decoding of PAC codes must polarize, meaning that the lower bounds increase exponentially with $N/2$ after one step of polarization \cite{moradi2023application}.

For a low complexity sequential decoding of convolutional codes, the number of data bits must be less than $NR_0(W)$.
For this scheme, it follows that at the first level of polarization, the number of data bits in the first half of the data carrier vector $\mathbf{v}$ must be less than $\frac{N}{2}R_0(W^-)$ and the number of data bits in the second half must be less than $\frac{N}{2}R_0(W^+)$. 
This increases the possible number of data bits from $NR_0(W)$ to the average $\frac{N}{2}\left(R_0(W^-)+R_0(W^+)\right)$, where we note
$R_0(W^-)+R_0(W^+) \geq 2R_0(W)$.
Based on this polarization, we propose a rate profiling technique along with an example in the following section.

\section{Search-Constrained Optimization Algorithm}\label{sec: optimization alg}
In this section, we detail our proposed method to construct a rate profile for a PAC$(256, 128)$ code using the polarization of the cutoff rate. 
The method is applicable to any code length and code rate. 
In the numerical results presented in this paper, we examine the performance of PAC$(256, 128)$ and PAC$(64, 32)$ codes.

\subsection{Optimization of Rate Profile}
The relation between the cutoff rate and the signal-to-noise ratio (SNR) is \cite[p.~303]{jacobs}
\begin{equation}
    R_0 = 1 - \log_2\left(\frac{2}{1+e^{-2E_b/N_0}}\right).
\end{equation}
To construct a conventional $(N, K)$ convolutional code, one should choose a design SNR, $\text{SNR}_d$, value such that the resulting cutoff rate $R_0$ is greater than $K/N$.

Consider the design SNR value to be $4~$dB.
At this SNR, the cutoff rate for a rate $R = 0.5$ code is $R_0(W) =0.513$. 
Therefore, for conventional convolutional codes, if a sequential decoder is to have a low computational complexity for SNR values greater than $4~$dB, the maximum number of bits that can be communicated must be less than $NR_0(W) \approx 131$.  
Based on the polarization of the computational cutoff rate, $K^- \leq \frac{N}{2}R_0(W^-) \approx 40$ and $K^+ \leq \frac{N}{2}R_0(W^+) \approx 100$.
Note that $100+40 = 140 > NR_0(W) \approx 131$, indicating that the coding rate can be improved after only one step of polarization. 
This now sets a constraint on the code rate profile in the sense that the first half of the profile should have no more than $40$ data bits, and the second half should contain no more than $100$ data bits. 
Note that polarization preserves capacity, but after one step of polarization, $R_0(W^-) = 0.3174$ and $R_0(W^+) = 0.7844$, yielding an average of $0.5509$, improved in comparison to $R_0(W) =0.513$.

Using the conventional superscript notation for polarization, we find that the second step of polarization yields $K^{--} \leq \frac{N}{4}R_0(W^{--}) \approx 8$, which represents the number of data bits in the first quarter of the coding rate profile. 
Similarly, $K^{-+} \approx 35$, $K^{+-} \approx 41$, and $K^{++} \approx 61$.
Note that $K^{--}+K^{-+}+K^{+-}+K^{++} \approx 145 > K^- + K^+ \approx 140$, which corresponds to an increase in the coding rate at the second step of polarization. 
Now, we are able to construct a rate $R= 145/256$ PAC code with a block length of $N=256$ that has a low computational complexity by assuming that only $8$ of the first $64$ bits can contain information bits. 
The second, third, and fourth $64$-bit locations should then include $35, 41$, and $61$ information bits, respectively. 
This establishes a constraint on the coding rate profile.

To construct a PAC$(256, 128)$ code, we consider the rate profile constraints imposed by the four-step cutoff rate polarization (the number of polarization steps is a parameter of the algorithm). 
In this case, $K^{----} + K^{---+} + \cdots + K^{++++} \approx 151$.
Based on the requisites provided by the fourth step of polarization, $128$ out of $151$ data bit locations should be selected. 
Compared to the selection from $256$ bit positions, this reduces the search space by a factor of nearly $6\times10^{48}$ times.
In the next section, we detail an evolutionary algorithm for this purpose.
Given these constraints, the search area is drastically decreased, resulting in much faster convergence than selecting $128$ data bit locations from $256$ places.

\subsection{Proposed Evolutionary Algorithm}
The fitness function that we use for a given SNR value is the maximum likelihood (ML) decoding upper bound value (MLUBV) on the error probability defined in \cite[p.~88]{viterbi2013principles}.
We obtain the weight distribution for the fitness function similar to \cite{li2012adaptive, tonnellier2021systematic}. 
In Algorithm \ref{alg1}, for a given rate profile, PAC code, list size $L$, and SNR value, we estimate the MLUBV, the code minimum distance $d$, and $A_d$, the number of codewords with the weight equal to $d$.
We refer to \cite{li2012adaptive} for details regarding obtaining weight distribution from list decoding.

\begin{algorithm}[htpb]\label{alg1}
\DontPrintSemicolon
\SetNoFillComment
\caption{$\textsc{FitnessFunction}$}
\KwIn{$\mathcal{A}, \mathbf{G} \triangleq \mathbf{T}\mathbf{F}^{\otimes n}, L, \text{SNR}$}
\KwOut{$\text{MLUBV}, d, A_d$}
$\hat{\mathbf{V}} \leftarrow \textit{ListDecoding}[\textbf{0}, \mathcal{A}, L]$\;
$\hat{\mathbf{X}} \leftarrow \hat{\mathbf{V}}\mathbf{G}$\;
$\mathbf{W}  \leftarrow \textit{sum}(\hat{\mathbf{X}}, 2)$\;
$(\text{MLUBV}, d, A_d)  \xleftarrow{\text{\cite[p.~88]{viterbi2013principles}}} [\mathbf{W}$, \text{SNR}]\;
\Return $[\text{MLUBV}, d, A_d]$\;
\end{algorithm}

We assume that the $\mathbf{0}$ vector is transmitted over a channel with no noise.
In Line 1 of the algorithm, using list decoding, we obtain the matrix $\hat{\mathbf{V}}$ of size $L$ by $N$ whose rows are the estimates of the data carrier vector $\mathbf{v}$.
We refer to \cite{yao2021list} for details regarding list decoding of PAC codes.
In Line 2, we obtain the matrix $\hat{\mathbf{X}}$ with rows containing estimates of the codeword vector $\mathbf{x}$.
The column-wise summation of the matrix $\hat{\mathbf{X}}$ yields the codeword weights, which, ideally, are closest to the transmitted $\mathbf{0}$ codeword.
Then, we calculate $d$, the estimated minimum distance of the codewords, and $A_{d}$, the estimations of the number of codewords with a weight equal to $d$. 
Finally, using $d$ and $A_{d}$, we calculate the fitness function of the rate profile $\mathcal{A}$ and return these values.

\begin{algorithm}[htpb]\label{Alg2}
\DontPrintSemicolon
\SetNoFillComment
\caption{$\textsc{DRPO}$}
\KwIn{$\mathbf{L}, [\mathcal{A}], d_{\text{best}}, A_{d_{\text{best}}}, \alpha, \mathbf{G}, \text{SNR}, \text{SNR}_d, p, maxiter$}
\KwOut{$\mathcal{A}$}
\For{$iter \leftarrow 1$ \KwTo maxiter}{

\For{$i \leftarrow 1$ \KwTo $PopSize$}{
\tcp*[h]{\scriptsize{for loop may be performed in parallel.}}\;
$[\text{MLUBV}_i, d_i, A_{d_i}] \leftarrow  \textsc{FitnessFunction}(\mathcal{A}_i, \mathbf{G}, L_1, \text{SNR})$\;
$l  \leftarrow 2$\;
\While{$(d_i > d_{\text{best}})$ or $($if $d_i = d_{\text{best}}$, $A_{d_i} \leq \alpha A_{d_{\text{best}}} )$}{
    $[\text{MLUBV}_i, d_i, A_{d_i}] \leftarrow  \textsc{FitnessFunction}(\mathcal{A}_i, \mathbf{G}, L_l, \text{SNR})$\;
    $l  \leftarrow l+1$; \tcp*[h]{\scriptsize{increase list size.}}
    
}
\If{$(d_i < d_{\text{best}})$ or $($$d_i = d_{\text{best}}$ and $A_{d, i} > \alpha A_{d_{\text{best}}} )$}{
        $\text{MLUBV}_i \leftarrow 1$\;
    }
}
$d_{\text{best}}  \leftarrow \max_{i= 1,2,\ldots, PopSize}(d_i)$\;
$A_{d_{\text{best}}} \leftarrow \min_{i= 1,2,\ldots, PopSize}\{A_{d_i} ~|~~ d_i = d_{\text{best}} \}$\;
$[\mathcal{A}]  \leftarrow \textsc{GenAlg}([\mathcal{A}], p, \text{SNR}_d)$\;
}
\Return $\mathcal{A}$: the rate profile corresponding to $A_{d_{\text{best}}}$ \;
\end{algorithm}

We now propose a dynamic evolutionary algorithm named dynamic rate profile optimization (DRPO) to optimize the rate profile for a PAC$(N, K)$ code. 
The steps are given in Algorithm \ref{Alg2}.
Input $\mathbf{L} = (L_1, L_2, \ldots, L_{\Lambda})$ is a vector whose elements $L_l$, $l = 1, 2, \ldots, \Lambda$, determine the list sizes for list decoding.
As its rows, the input matrix $[\mathcal{A}]$ contains $PopSize$ rate profiles with a minimum distance of at least $d_{\text{best}}$ (in our simulations we used $PopSize= 100$).
Using the guessing technique \cite{moradi2023application} and the RM rate profile, a random population of rate profiles with a minimum distance of $d_{\text{best}}$ can be generated.
If the $A_{d_i}$ values corresponding to the rate profiles $\mathcal{A}_i$ in $[\mathcal{A}]$ are known, $A_{d_{\text{best}}}$ can be set to the minimum of these values; otherwise $A_{d_{\text{best}}} = \infty$.
Input $\alpha$ is a constant number greater than $1$ (in this paper, we used $\alpha = 1.2$).
Input $\text{SNR}_d$ is the design SNR.

In Lines 3 and 6, we obtain and store $d_i$, $A_{d_i}$, and $\text{MLUBV}_i$ alongside the $\mathcal{A}_i$ rate profile\footnote{We note that alternative metrics, such as bit error rate (BER) could be used for the fitness function, as well as modification of the code structure, e.g., to systematic PAC codes. 
The algorithm would operate in the same way for these cases, with the exception of the modified fitness function and code structure, respectively.}.
In Line 15, the genetic algorithm \textsc{GenAlg} is based on the one provided in \cite{elkelesh2019decoder}, where the population is extended using only selection and mutation.
For the selection method, $p$ percent of the rate profiles with the best fitness function values (elite rate profiles) advance to the subsequent iteration.
For the mutation method, we randomly select an elite rate profile and have a mutation on a randomly chosen bit position.
In this paper, we used $p = 5$.
After mutation, the remaining bit positions will be randomly flipped until the code has a rate of $K/N$ and the constraint imposed by the polarization tree is met
\footnote{The complete MATLAB code based on the pseudocode presented in this paper is available at https://github.com/moradi-coding/.}.

In existing algorithms, the \textsc{FitnessFunction} procedure employs a large list size for all candidates. 
However, this algorithm takes advantage of the fact that if a rate profile for a very small list size produces a poor $d_{\min}$, or a good $d_{\min}$ with a very large $A_{d_{\min}}$, we may immediately determine that the rate profile is bad (compared to the other considered rate profiles) and there is no need to consider larger list size to obtain its weight distribution.
The numerical results in the following section demonstrate that this dynamic list size significantly accelerates the evolutionary algorithm.

We also note that, in our proposed algorithm, the constraint imposed by the polarization tree adds more variance to the rate profile and provides a benefit where the genetic algorithm randomly searches a larger neighborhood of the elite population.
The numerical results in the next section demonstrate that the randomness of the mutation is sufficient to discover a suitable profile, thereby simplifying the evolutionary algorithm.

Finally, as the numerical results in Section \ref{sec: Numerical} demonstrate, only a few of the $PopSize =100$ rate profiles necessitate SCL decoding with a large list size.
Due to this decreased complexity, our algorithm is able to calculate the fitness function of all $PopSize$ rate profiles in parallel, unlike \cite{tonnellier2021systematic}.
These improvements significantly increase the speed of the optimization such that even an ordinary computer can obtain all fitness values in parallel.

\section{Numerical Results}\label{sec: Numerical}

\subsection{PAC(256, 128) code:}

Table. \ref{tab: N256_K128_barplot} displays the number of fitness function evaluations performed by the proposed evolutionary algorithm for the PAC$(256, 128)$ code over the course of $1000$ iterations for the list size vector $\mathbf{L} = (100, 200, 500, 1000, 5000, 30000)$.
The constrained search space of the data bits is specified based on the fourth step of the cutoff rate polarization with $\text{SNR}_d =4~$dB\footnote{Note that the larger SNR$_d$ value only enlarges the search space, so the FER performance would only be improved at the cost of (asymptotically [9]) infinite ANV values for the SNR values below the larger SNR$_d$ value (SNR$_d = \infty$ in [10]).}.
Out of the $100,000$ considered rate profiles (recall, $PopSize = 100$), the algorithm observes repeated rate profiles $5910$ times.
This is because of the small rate profile space and small block length.
The algorithm executed the fitness function $94090$ times with a list size of $100$ and saved $45263$ calls to the fitness function for lists with a larger size.
For the extreme value, a list size of $L = 30,000$, the fitness function was executed only $7553$ times, which accounts for approximately 8$\%$ of the instances in which it encounters new rate profiles.
In this example, we initialized the algorithm with the values $d_{\text{best}} = 18$ and $A_{d_{\text{best}}} = \infty$.
In addition, to construct the input matrix $[\mathcal{A}]$, we randomly picked $128$ bit positions from the rate profile of RM$(256, 163)$ that satisfies constraints by the cutoff rate polarization at the fourth step.

\begin{table}[h]
\setlength{\tabcolsep}{3.5pt}
\centering
\caption{Number of the fitness function evaluations of a PAC$(256, 128)$ code for 1000 iterations of the optimization.}
\begin{tabular}{c||ccccccc}
List size & 0 & 100 & 200 & 500 & 1000 & 5000 & 30000 \\
\hline \hline
\#evaluations & 100000 & 94090 & 48827 & 28544 & 18779 & 13133 & 7553 \\
\end{tabular}\label{tab: N256_K128_barplot}
\end{table}

\begin{figure}[h] 
\centering
	\includegraphics [width = \columnwidth]{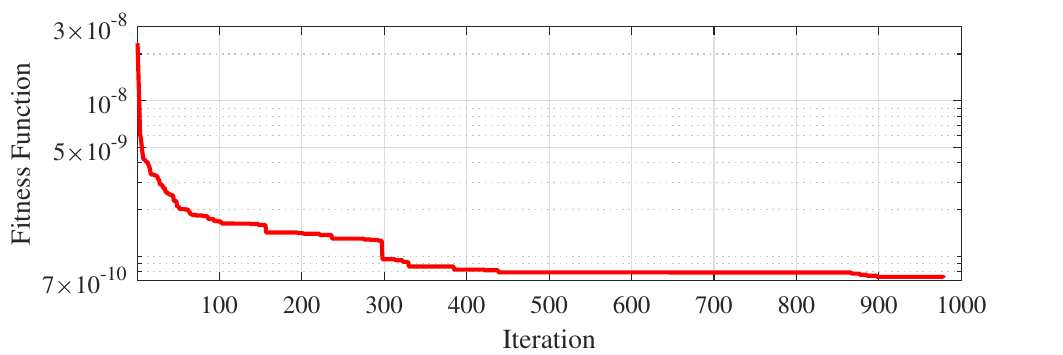}
	\caption{Fitness function values of a PAC$(256, 128)$ code for different iterations.} 
	\label{fig: N256_K128_CostFun}
\end{figure}

Fig. \ref{fig: N256_K128_CostFun} depicts the best fitness function values at each iteration of the algorithm (assessed at an SNR value of $4~$dB).
As shown in this figure, the value of the fitness function decreases with iterations and drops sharply at the $298$th iteration, after which the decrease continues but not significantly.
At the $298$th iteration, the minimum distance is $d_{\text{best}} = 20$, with $A_{20} = 158$, $A_{22} = 0$, and $A_{24} = 988$.
The last improvement is in the $979$th iteration, and out of the $5$ elite rate profiles, the best one has a minimum distance of $20$, with $A_{20} = 126$, $A_{22} = 0$, and $A_{24} = 711$.
Note that the optimization of \cite{tonnellier2021systematic} results in a minimum distance of $20$, with $A_{20} = 430$, $A_{22} = 68$, and $A_{24} = 6709$.
For this comparison, to be the same as \cite{tonnellier2021systematic}, we used a list size of $262144$.

\begin{figure}[h] 
\centering
	\includegraphics [width = \columnwidth]{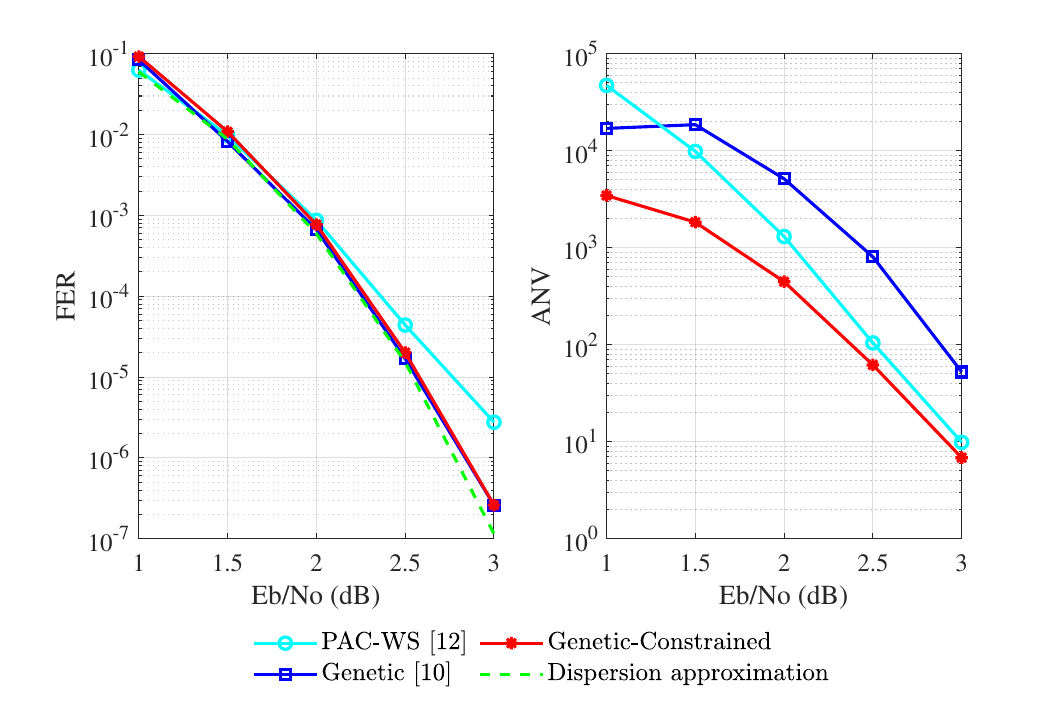}
	\caption{Performance comparison of PAC$(256, 128)$ codes.} 
	\label{fig: N256_K128_GenSysvGenTame}
\end{figure}

Fig. \ref{fig: N256_K128_GenSysvGenTame} displays a comparison of the error-correction performance of the PAC$(256, 128)$ code when its rate profile is constructed using a standard genetic algorithm as described in \cite{tonnellier2021systematic} versus when it is constructed using our proposed rate profile construction. 
The results indicate that our proposed rate profile construction performs approximately the same in terms of error-correction performance.
This figure also illustrates the corresponding average number of Fano decoding visits per decoded bit.
The ANV value for our proposed rate profile is around $356$ at an SNR value of $2~$dB, but, using a conventional genetic algorithm, it is approximately $5137$, representing an improvement of approximately $92$\%. 
This figure also includes a plot of the Fano decoding performance for the PAC code from \cite{liu2022weighted}. 
In this paper, the threshold spacing parameter $\Delta = 2$ is used in all simulations of Fano decoding.

\begin{table*}[ht] 
\centering
\caption{Rate profiles.}
\renewcommand{\arraystretch}{1.2}
\begin{tabular}{ccc}
\hline
$(N,K)$                      & Design SNR (dB) &  Rate profile $\mathcal{A}$ (hexadecimal)                                                 \\ \hline
$(256,128)$ & 4    & 0000000400808745000A1737175FBBEF011A177F957F7AFF1EFFD6BFE77A79D7 \\ \hline 

$(64,32)$   & 5        & 000A467F9CCE937F                                                 \\ \hline 

$(128,105)$   & $7$        & 173F37BF97FF7FEF177D7FFF7FFFFFFF                                                 \\ \hline 
$(64,51)$   & $7$        & D5DF7BEFD7DFBF77                                                 \\ \hline 
\end{tabular}\label{tab: rate profile}
\end{table*}



\subsection{PAC(64, 32) code:}
Table. \ref{tab: N64_K32_barplot} displays the number of fitness function evaluations performed by the proposed evolutionary algorithm for the PAC$(64, 32)$ code over the course of $200$ iterations for the list size vector $\mathbf{L} = (100, 200, 1000, 50000)$.
The data bits constrained search space is specified using the third step of the cutoff rate polarization and $\text{SNR}_d =5~$dB.
As this table demonstrates, the algorithm needs to invoke the fitness function with a large list size of $L = 50,000$ only $215$ times, or nearly 1$\%$ of the times it encounters new rate profiles.
$100$ of the $215$ calls are for the first iteration and could be reduced by finding a better rate profile in order to reduce to $A_{d_{\text{best}}}$.
We used $d_{\text{best}} = 8$ and $A_{d_{\text{best}}} = \infty$ as the initialization of the algorithm.
The rate profiles of the constructed codes are displayed in Table \ref{tab: rate profile}.
They are reported in hexadecimal format, with zeros representing the frozen bits.

\begin{table}[h]
\setlength{\tabcolsep}{5pt}
\centering
\caption{Number of the fitness function evaluations of a PAC$(64, 32)$ code for a 200 number of iterations.}
\begin{tabular}{c||ccccc}
List size & 0 & 100 & 200 & 1000 & 50000 \\
\hline \hline
\#evaluations & 20000 & 12623 & 396 & 215 & 215 \\
\end{tabular}\label{tab: N64_K32_barplot}
\end{table}

\begin{figure}[h] 
\centering
	\includegraphics [width = \columnwidth]{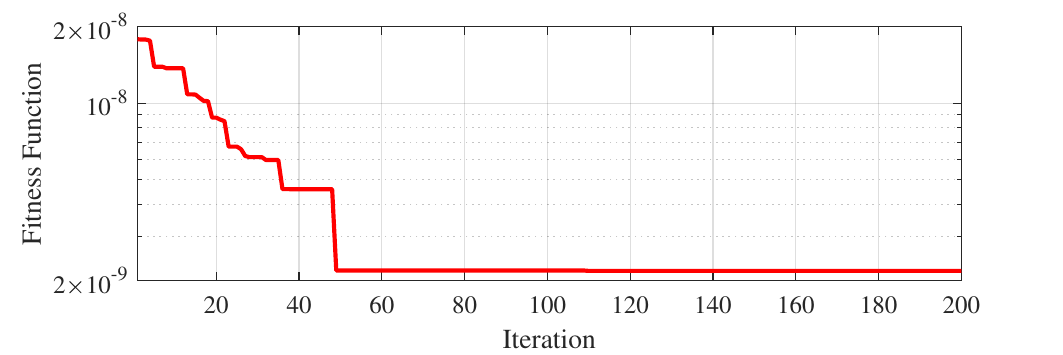}
	\caption{Fitness function value of a PAC$(64, 32)$ code for different iterations.} 
	\label{fig: N64_K32_CostFun}
\end{figure}

Fig. \ref{fig: N64_K32_CostFun} depicts the corresponding fitness function values calculated at an SNR of $7~$dB. 
The minimum distance of the code is $8$ until the $48$th iteration, improving to $10$ at the $49$th iteration, which results in a significant improvement of the fitness function value.
After the $49$th iteration, the fitness function value remains nearly unchanged, and after the $110$th iteration, it experiences a minor improvement before remaining unchanged until the $200$th iteration.

\begin{figure}[h] 
\centering
	\includegraphics [width = \columnwidth]{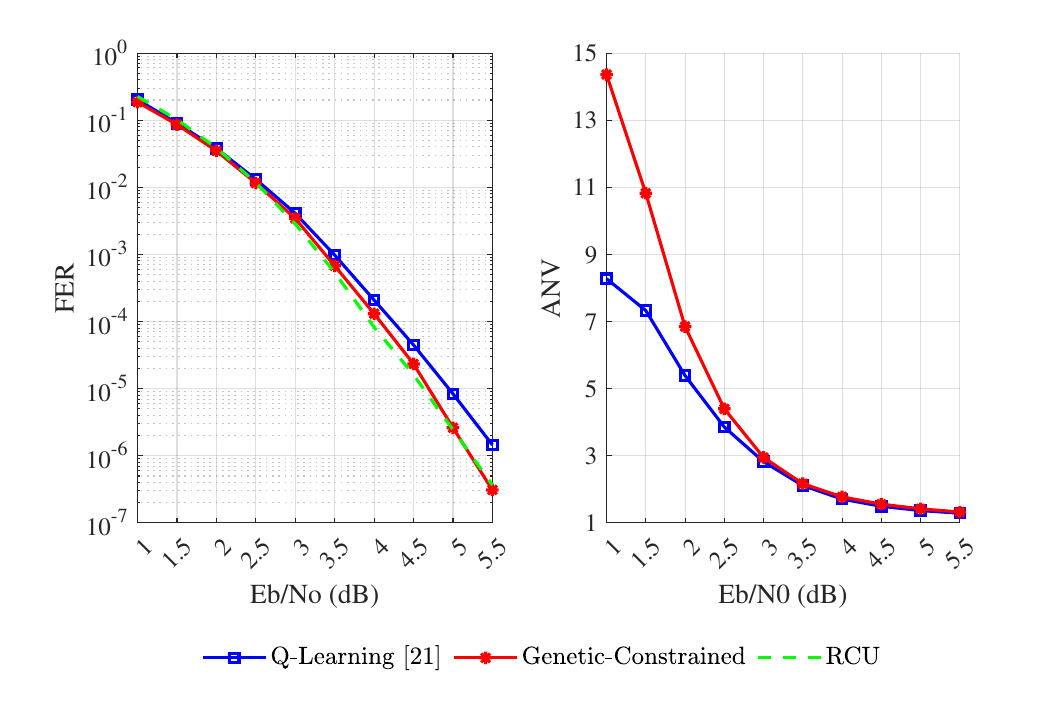}
	\caption{Performance comparison of PAC$(64, 32)$ codes.} 
	\label{fig: N64_K32_GenGuessing}
\end{figure}

Fig. \ref{fig: N64_K32_GenGuessing} compares the performance of the PAC$(64, 32)$ code when its rate profile is constructed using a Q-Learning algorithm as described in \cite{mishra2022modified} versus when its rate profile is constructed using our proposed method. 
The results indicate that, unlike Q-learning, our method enables the Fano algorithm to achieve the random coding union (RCU) bound with comparable computational complexity. 
For both results, the decoder has the same $\Delta = 2$.

\subsection{PAC(128, 105) code:}

\begin{table}[h]
\setlength{\tabcolsep}{5pt}
\centering
\caption{Number of the fitness function evaluations of a PAC$(128, 105)$ code for a 1000 number of iterations.}
\begin{tabular}{c||ccccc}
List size & 0 & 100 & 1000 & 2000 & 200000 \\
\hline \hline
\#evaluations & 100000 & 13739 & 5843 & 2841 & 2836 \\
\end{tabular}\label{tab: N128_K105_barplot}
\end{table}

Table \ref{tab: N128_K105_barplot} presents the number of fitness function evaluations conducted by the proposed evolutionary algorithm for the high-rate PAC$(128, 105)$ code across $1000$ iterations, considering list size vector $\mathbf{L} = (100, 1000, 2000, 200000)$.
The search space for the data bits is constrained using the third step of the cutoff rate polarization, with $\text{SNR}_d = 7~$dB. 
As depicted in the table, the algorithm requires invoking the fitness function with a large list size of $L = 200,000$ in nearly 2.8$\%$ of the encounters with new rate profiles.
We initialized the algorithm with $d_{\text{best}} = 6$ and $A_{d_{\text{best}}} = 1308$, derived from the rate profile described in \cite{moradi2023application}. 
The remaining $99$ initial rate profiles are random profiles that satisfy the constraints imposed by $\text{SNR}_d = 7~$dB.

\begin{figure}[h] 
\centering
	\includegraphics [width = \columnwidth]{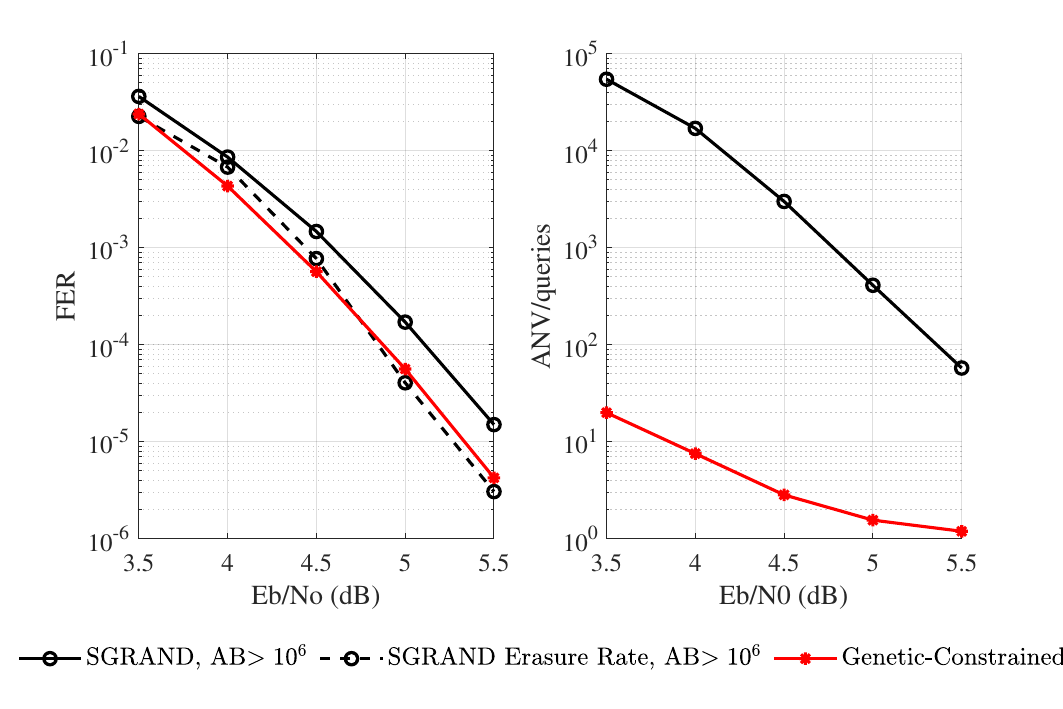}
	\caption{Performance comparison of $(128, 105)$ codes.} 
	\label{fig: N128_K105_GenGuessing}
\end{figure}

Fig. \ref{fig: N128_K105_GenGuessing} presents a comparison between the performance of Soft Guessing Random Additive Noise Decoding (SGRAND) for a $(128, 105)$ CA-Polar code, employing an ABandonment (AB) threshold of $10^6$ queries, as described in \cite{solomon2020soft}, and Fano decoding applied to our proposed PAC$(128, 105)$ code. 
The PAC code exhibits a coding gain of approximately $0.25~$dB in FER of $10^{-6}$. 
This figure also depicts the average number of queries required by SGRAND and the ANV values required by Fano decoding.
During each query, the SGRAND algorithm verifies if the decoded codeword is valid, achievable through vector and matrix multiplication. 
It is noteworthy that the GRAND algorithms have the capability to execute multiple queries simultaneously, albeit with higher memory usage, and mostly suited for high-rate codes. 
On the other hand, the Fano algorithm requires very little memory and, with the selection of an appropriate rate profile for the PAC code, can achieve very low computational complexity.
Additionally, the figure illustrates the erasure rate concerning the abandonment threshold of $10^6$ codebook queries. 
If the algorithm fails to discover a valid codeword within $10^6$ queries, those instances are considered as correctly decoded codewords for the erasure rate calculation. 
Hence, this serves as an upper bound on the performance of SGRAND.

\begin{table}[h]
\setlength{\tabcolsep}{5pt}
\centering
\caption{Number of the fitness function evaluations of a PAC$(64, 51)$ code for a 1000 number of iterations.}
\begin{tabular}{c||ccccc}
List size & 0 & 100 & 1000 & 10000 & 200000 \\
\hline \hline
\#evaluations & 100000 & 17361 & 4505 & 1690 & 390 \\
\end{tabular}\label{tab: N64_K51_barplot}
\end{table}

\subsection{PAC(64, 51) code:}

Table \ref{tab: N64_K51_barplot} displays the number of fitness function evaluations for the high-rate PAC$(64, 51)$ code over $1000$ iterations, considering a list size vector $\mathbf{L} = (100, 1000, 10000, 200000)$. 
The search space is constrained using the third step of the cutoff rate polarization and with $\text{SNR}_d$ set at $7$ dB.
For initializing the algorithm, we used $d_{\text{best}} = 4$ and $A_{d_{\text{best}}} = 132$, which corresponds to the rate profile constructed from \cite{moradi2023application}. 
The remaining $99$ initial rate profiles are randomly generated profiles that satisfy the constraint set by $\text{SNR}_d = 7$ dB.

\begin{figure}[h] 
\centering
	\includegraphics [width = \columnwidth]{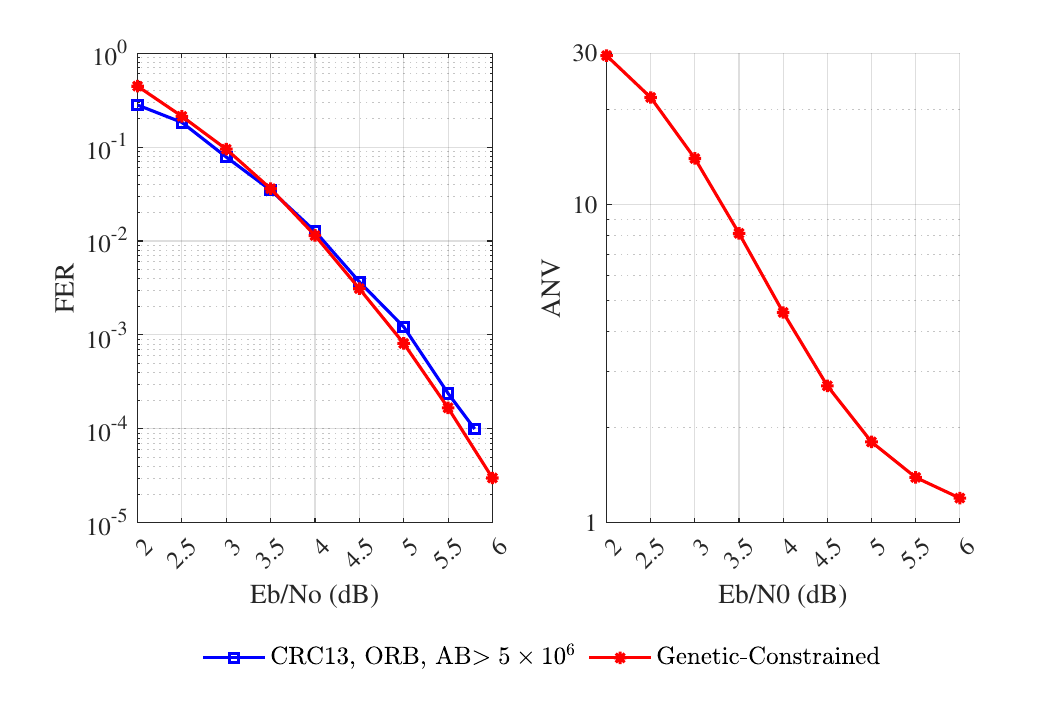}
	\caption{Performance comparison of $(64, 51)$ codes.} 
	\label{fig: N64_K51_GenGuessing}
\end{figure}

Fig. \ref{fig: N64_K51_GenGuessing} compares the performance of the Ordered Reliability Bits (ORB) GRAND for a $(64, 51)$ CRC code with an abandonment (AB) threshold of $5\times10^6$ queries from \cite{an2021crc} and Fano decoding applied to our proposed PAC$(64, 51)$ code.

\section{Conclusion}\label{sec: conclusion}
In this paper, we proposed a novel evolutionary-based rate profile design method that improves the code error-correction performance under the search space constraint imposed by the polarization of the computational cutoff rate.
Using a dynamic SCL decoding evaluation in the fitness function makes our algorithm significantly more efficient than those in the literature.
In addition, the computational complexity versus error-correction performance tradeoff in our paper is significantly improved compared to previous works.
These optimized short codes perform close to the random coding bound with low complexity, and thus, we believe are good candidates for Ultra-Reliable Low Latency Communications (URLLC) applications like autonomous driving, factory automation, and virtual/augmented reality.

\bibliographystyle{IEEEtran}
\bibliography{bibliography}

\begin{thebibliography}{10}
\providecommand{\url}[1]{#1}
\csname url@samestyle\endcsname
\providecommand{\newblock}{\relax}
\providecommand{\bibinfo}[2]{#2}
\providecommand{\BIBentrySTDinterwordspacing}{\spaceskip=0pt\relax}
\providecommand{\BIBentryALTinterwordstretchfactor}{4}
\providecommand{\BIBentryALTinterwordspacing}{\spaceskip=\fontdimen2\font plus
\BIBentryALTinterwordstretchfactor\fontdimen3\font minus \fontdimen4\font\relax}
\providecommand{\BIBforeignlanguage}[2]{{%
\expandafter\ifx\csname l@#1\endcsname\relax
\typeout{** WARNING: IEEEtran.bst: No hyphenation pattern has been}%
\typeout{** loaded for the language `#1'. Using the pattern for}%
\typeout{** the default language instead.}%
\else
\language=\csname l@#1\endcsname
\fi
#2}}
\providecommand{\BIBdecl}{\relax}
\BIBdecl

\bibitem{polyanskiy2010channel}
Y.~Polyanskiy, H.~V. Poor, and S.~Verd{\'u}, ``Channel coding rate in the finite blocklength regime,'' \emph{IEEE Transactions on Information Theory}, vol.~56, no.~5, pp. 2307--2359, 2010.

\bibitem{arikan2019sequential}
E.~Ar{\i}kan, ``From sequential decoding to channel polarization and back again,'' \emph{arXiv preprint arXiv:1908.09594}, 2019.

\bibitem{moradi2020performance}
M.~Moradi, A.~Mozammel, K.~Qin, and E.~Ar{\i}kan, ``Performance and complexity of sequential decoding of {PAC} codes,'' \emph{arXiv preprint arXiv:2012.04990}, 2020.

\bibitem{fano1963heuristic}
R.~Fano, ``A heuristic discussion of probabilistic decoding,'' \emph{IEEE Transactions on Information Theory}, vol.~9, no.~2, pp. 64--74, 1963.

\bibitem{tal2015list}
I.~Tal and A.~Vardy, ``List decoding of polar codes,'' \emph{IEEE Transactions on Information Theory}, vol.~61, no.~5, pp. 2213--2226, 2015.

\bibitem{li2019pre}
B.~Li, H.~Zhang, and J.~Gu, ``On pre-transformed polar codes,'' \emph{arXiv preprint arXiv:1912.06359}, 2019.

\bibitem{moradi2023minimum}
M.~Moradi, ``Polarization-adjusted convolutional ({PAC}) codes as a concatenation of inner cyclic and outer polar-and {R}eed-{M}uller-like codes,'' \emph{Finite Fields and Their Applications}, vol.~93, p. 102321, 2024.

\bibitem{yao2023deterministic}
H.~Yao, A.~Fazeli, and A.~Vardy, ``A deterministic algorithm for computing the weight distribution of polar code,'' \emph{IEEE Transactions on Information Theory}, 2023.

\bibitem{moradi2023application}
M.~Moradi, ``Application of guessing to sequential decoding of polarization-adjusted convolutional ({PAC}) codes,'' \emph{IEEE Transactions on Communications}, 2023.

\bibitem{tonnellier2021systematic}
T.~Tonnellier and W.~J. Gross, ``On systematic polarization-adjusted convolutional ({PAC}) codes,'' \emph{IEEE Communications Letters}, vol.~25, no.~7, pp. 2128--2132, 2021.

\bibitem{moradi2021monte}
M.~Moradi and A.~Mozammel, ``A {M}onte-{C}arlo based construction of polarization-adjusted convolutional ({PAC}) codes,'' \emph{arXiv preprint arXiv:2106.08118}, 2021.

\bibitem{liu2022weighted}
W.~Liu, L.~Chen, and X.~Liu, ``A weighted sum based construction of {PAC} codes,'' \emph{IEEE Communications Letters}, 2022.

\bibitem{arikan2006channel}
E.~Ar{\i}kan, ``Channel combining and splitting for cutoff rate improvement,'' \emph{IEEE Transactions on Information Theory}, vol.~52, no.~2, pp. 628--639, 2006.

\bibitem{moradi2021sequential}
M.~Moradi, ``On sequential decoding metric function of polarization-adjusted convolutional ({PAC}) codes,'' \emph{IEEE Transactions on Communications}, vol.~69, no.~12, pp. 7913--7922, 2021.

\bibitem{arikan1996inequality}
E.~Ar{\i}kan, ``An inequality on guessing and its application to sequential decoding,'' \emph{IEEE Transactions on Information Theory}, vol.~42, no.~1, pp. 99--105, 1996.

\bibitem{jacobs}
I.~M. Jacobs and J.~Wozencraft, \emph{Principles of {Communication} {Engineering}.}, 1965.

\bibitem{viterbi2013principles}
A.~J. Viterbi and J.~K. Omura, \emph{Principles of digital communication and coding}.\hskip 1em plus 0.5em minus 0.4em\relax New York: McGraw-Hill, 1979.

\bibitem{li2012adaptive}
B.~Li, H.~Shen, and D.~Tse, ``An adaptive successive cancellation list decoder for polar codes with cyclic redundancy check,'' \emph{IEEE communications letters}, vol.~16, no.~12, pp. 2044--2047, 2012.

\bibitem{yao2021list}
H.~Yao, A.~Fazeli, and A.~Vardy, ``List decoding of ar{\i}kan’s {PAC} codes,'' \emph{Entropy}, vol.~23, no.~7, p. 841, 2021.

\bibitem{elkelesh2019decoder}
A.~Elkelesh, M.~Ebada, S.~Cammerer, and S.~Ten~Brink, ``Decoder-tailored polar code design using the genetic algorithm,'' \emph{IEEE Transactions on Communications}, vol.~67, no.~7, pp. 4521--4534, 2019.

\bibitem{mishra2022modified}
S.~K. Mishra, D.~Katyal, and S.~A. Ganapathi, ``A modified {Q}-learning algorithm for rate-profiling of polarization adjusted convolutional ({PAC}) codes,'' \emph{2022 IEEE Wireless Communications and Networking Conference (WCNC)}, pp. 2363--2368, 2022.

\bibitem{solomon2020soft}
A.~Solomon, K.~R. Duffy, and M.~M{\'e}dard, ``Soft maximum likelihood decoding using grand,'' in \emph{ICC 2020-2020 IEEE International Conference on Communications (ICC)}.\hskip 1em plus 0.5em minus 0.4em\relax IEEE, 2020, pp. 1--6.

\bibitem{an2021crc}
W.~An, M.~M{\'e}dard, and K.~R. Duffy, ``Crc codes as error correction codes,'' in \emph{ICC 2021-IEEE International Conference on Communications}.\hskip 1em plus 0.5em minus 0.4em\relax IEEE, 2021, pp. 1--6.

\end{thebibliography}

\end{document}